%% file: elsarticle-template-harv.tex
\documentclass[preprint,12pt,authoryear]{elsarticle}

\usepackage[ruled,vlined]{algorithm2e}
\usepackage{amssymb}
\usepackage{amsthm}
\usepackage{amsmath}
\usepackage{multirow}
\usepackage{caption}
\usepackage{graphicx}

\journal{Journal of Financial Economics}

\begin{document}

\begin{frontmatter}



\title{Multivariate Pair Trading by Volatility \& Model Adaption Trade-off}


\author[inst1]{Chenyanzi Yu\footnotemark[1]}
\author[inst2]{Tianyang Xie\footnotemark[1]}

\affiliation[inst1]{organization={Department of Mathematics, University of Chicago},
            country={US}}

\affiliation[inst2]{organization={Eller College of Management, University of Arizona},
            country={US}}
            
\begin{abstract}
Pair trading is one of the most discussed topics among financial researches. Despite a growing base of work, portfolio management for multivariate time series is rarely discussed. On the other hand, most researches focus on refining strategy rules instead of finding the optimal portfolio weight. In this paper, we brought up a simple yet profitable strategy called Volatility \& Model Adaption Trade-off (VMAT) to leverage the issues. Experiment studies show its superior profit performance over baselines.

\end{abstract}


\end{frontmatter}


\footnotetext[1]{Both authors share the equal contribution.}

\input{section_1}

\input{section_2}

\input{section_3}

\input{section_4}

\bibliographystyle{elsarticle-harv} 
\bibliography{cas-refs}

\appendix

\input{appendix_1}

\input{appendix_2}

\input{appendix_3}

\input{appendix_4}






\end{document}

%% file: section_1.tex
\section{Introduction}
\label{section_1}

Trading across equities is a long lasting research hotspot over time. The classical approach of using a simple two step method to determine the relationship between two time series \citep{Gatev2006} has been well established both theoretically and empirically \citep{Do2010}. Later, more refined correlation methods have been introduced \citep{Perlin2007, Chen2012}. Beside the simple algorithmic approaches, methods that assess the comovement between pairs by cointegration testing have come into spotlight \citep{Vidyamurthy2004}. For multivariate time series, generalized cointegration approaches have been introduced \citep{Galenko2007, Dunis2016}. In the domain of utilizing time series forecasting techniques, there is also the classical method that model the simple spread between the pairs of equities \citep{Elliott2005} as well as more complex framework based on modern machine learning methods \citep{Huck2010}.

Despite the growing base of work, there is only a relatively small amount of strategies designed for multivariate scenarios. On the other hand, while most methods concentrate on refining the strategy rules (i.e. when to participate and when to bail out), they put less discussion in determining the portfolio weight allocation. However, optimizing the portfolio weight is crucial for profiting.

To leverage the issues, we introduce our strategy framework called Volatility \& Model Adaption Trade-off (VMAT) in this paper. The framework first finds optimal portfolio weight for multivariate time series by solving a trade-off between AR model predictability and volatility of the portfolio process. Then, it determines participation by controlling the profiting probability in the trading period. Experiment studies show that VMAT has superior profit performance over the baselines including the traditional cointegration method especially when the length of trading period is limited. 

The paper is organized as the following: in section \ref{section_2}, we unify the problem formulation for the pair trading research, and then develop our framework by generalizing traditional baselines; in section \ref{section_3}, we study multiple empirical issues of VMAT based on real data experiments; in section \ref{section_4}, we summarize VMAT's advantages and further consideration.

%% file: section_2.tex
\section{Method}
\label{section_2}

\subsection{Problem Formulation}

Formulate equities as a multivariate time series $X_t \in \mathcal{R}_+^p$. A trading strategy of a limited trading period $d$ at time $t$ is characterized by three elements: 
\begin{itemize}
    \item Portfolio weight vector $w_t \in [-1,1]^p$.
    \item Signal of participation $\delta_t \in \{-1, 0, 1\}$.
    \item Time to bail $l_t \in \mathcal{N^+}$.
\end{itemize}

The strategy aims to maximize the log return of the portfolio.

\begin{align}\label{equation_1}
    \text{maximize}_{w_t, \delta_t, l_t} \quad \mathbf{E}_{X_{t+l_t}|X_{(-\infty):(t-1)}} & \{\delta_t w_t^T (log(X_{t+l_t}) - log(X_t))\} \notag \\
    & \text{s.t.} \quad |w_t|_1 = 1
\end{align}

For the simplicity of discussion, we employ a greedy approach for time to bail function $l_t$:

\begin{equation}
    \tilde {l_t} = min \{k \in \mathcal{N}^+: X_{t+k} > X_t\}
\end{equation}

It means that the strategy finishes the arbitrage as long as profit emerges. For the rest of the discussion in this paper, we focus on developing $w_t, \delta_t$.

\subsection{What's Beyond Baselines?}

Classical statistical arbitrage methods aim to form a long term equilibrium relationship between multivariate series \citep{Galenko2007}. The strategy tends to find abnormal portfolio value and then perform the arbitrage over a future time point where the value will revert to the equilibrium (a.k.a mean reverting). The strategy is formally called cointegration approach \citep{Krauss2015}.

Empirically, the strategy first finds a weight vector $w_t$ such that $y_t = w_t^T log X_t$ is stationary. Then the strategy selects a $\delta_t$ such that the profiting probability will be controlled in the following trading period. Due to the stationarity of the process, the probability of profiting in the trading period can be approximated by the profiting probability on the stationary distribution when $d \rightarrow \infty$. Additionally, by assuming the stationary distribution is asymptotically normal and we can approximate the joint profiting probability by the product of the single variate profiting probability, we can derive a $\delta_t$ that control the profiting probability over level $\alpha$:

\begin{align}
    \delta_{\alpha, t} &= \begin{cases}
    1 & y_t < \sigma^{Long}_{\alpha, t} \\
    0 & \text{else} \\
    -1 & y_t > \sigma^{Short}_{\alpha, t}
    \end{cases} \notag \\
    \text{where} \quad \sigma^{Short}_{\alpha, t} &= N_{(1 - \alpha)^{\frac{1}{d}}/2}\{\mathrm{E} (y^*_t),Var(y^*_t)\} \notag \\
    \quad \sigma^{Long}_{\alpha, t} &= - \sigma^{Short}_{\alpha, t} 
\end{align}

In practice, the mean and variance of asymptotic normal distribution is then estimated by the limited sample of formation period.

The method heavily relies on the assumption of the trading period $d$ approaches infinity, which is not the case in real life due to the customer demand or the risk of low liquidity. When $d$ is limited, the exact profiting probability can not be approximated.

A naturally question will be followed: what if we can employ time series techniques to model the stationary univariate time series, so that we can know better of the exact profiting probability? It's easy to see the plausibility. For example, if we assume the stationary univariate time series truly follows an AR process. Additionally, we can consistently estimate the coefficients of the correct AR structure. Then, by employing this additional modeling knowledge, we can formulate the exact profiting probability by comparing to the forecast intervals of AR model.

One step further, if we have a strong forecasting model that not only can fit stationary univariate time series, but also a larger class of series, do we still have to form a stationary process from the multivariate series? This question leads us to probe into the essence of profitability of statistical arbitrage strategies.

In our understanding, the profitability of statistical arbitrage strategies can be separated into two parts: predictability and volatility. While the volatility describes the scale of potential profit, the predictability represents the likelihood of profiting. 

The traditional cointegration approach focus on raising predictability of the portfolio series by deriving stationarity, but ignored profitability conveyed through the volatility of the series, thus traditional cointegration approach is sub-optimal.

On the other hand, one can raise the volatility of a formed series to obtain larger scale of profiting margin, but the series may become highly unpredictable. Therefore, increasing in the overall profit will not be ensured.

In the next section, we will introduce a framework called Volatility \& Model Adaption Trade-off (VMAT) to entackle the above issue. The VMAT framework employs AR model for univariate time series and adopts the concept of profiting probability control from the traditional cointegration approach. It extends the profitability from predictability only to an optimal trade-off between volatility and predictability.

\subsection{VMAT: Volatility \& Model Adaption Trade-off}

At each trading time $t$, VMAT finds $w_{t}$ to optimize the trade-off between predictability and volatility. 

\begin{align}
    \text{maximize}_{w_{t}} \quad \textbf{Predictability}(w_{t}^T log X_{t}) &+ \lambda * \textbf{Volatility}(w_{t}^T log X_{t}) \notag \\
    \text{s.t.} \quad |w_{t}|_1 &= 1 \notag
\end{align}

Note that $\lambda$ is a tuning parameter that controls the relative importance between volatility and predictability. While the volatility of an univariate time series is easy to estimate, it's hard to measure the predictability of the series on a given model framework. For simplicity of optimization, we utilize a simple AR model and measure the predictability by its squared error loss.

Then we rewrite the optimization as the following and solve $w_t$ independently at each trading time $t$:

\begin{align}
    \text{maximize}_{w_t, \beta_t} \quad - \quad \sum_{t'=t - L}^{t} (w_t^T log X_{t'} - w_t^T [log X_{(t'-p):(t'-1)}] \beta_t)^2 & \notag \\ 
    +\quad \lambda \sum_{t'=t - L}^{t} (w_t^T log X_{t'} &- w_t^T \overline{log X})^2 \notag \\
    \text{s.t.} \quad |w_t|_2 = 1 \quad \quad \quad \quad \quad \quad & 
\end{align}

In the above optimization, we take the last $L+1$ data in the history of time $t$ to train $w_t, \beta_t$ where $\beta_t$ is the coefficient vector for AR(p) model; $[log X_{(t'-p):(t'-1)}]$ is the $k \times p$ design matrix for AR(p) model where $k$ is the dimension for multivariate series; $L$ as a parameter is the length of the formation period; $\overline{log X}$ is the sample mean of $log X_{t'}$ over the formation period. Also note that the constraint is changed from L1 to L2 for easiness of the optimization.

The optimization conveys the idea of trade-off through the parameter $\lambda$. When $\lambda$ approaches infinity, the optimal $w_t$ becomes the weight vector $w_{t}^{MaxVar}$ that tries to maximize univariate series volatility. When $\lambda = 1$, the objective becomes the likelihood ratio statistic of the AR(p) model over a null model (i.e. sample mean). Therefore, the tuned $\lambda$ should be between 1 and infinity.

For solving the optimization, we can see that the partial optimal estimator $\hat \beta_t(w_t)$ is simply the OLS estimator for AR(p) model on the univariate time series $\{w_t log X_{t'}\}$. For the partial optimal estimator for $\hat w_t(\beta_t)$, we are solving a quadratic programming problem where $w_t$ is constrained on a unit sphere.

\begin{align}
    maximize_{w_t} \quad w_t^T K(\beta_t) w_t \notag \\
    \text{s.t.} \quad |w_t|_2 = 1
\end{align}

$K(\beta_t)$ is a $k \times k$ matrix defined by $\beta_t$ and the data $\{X_t'\}$ from the optimization. Solving the above quadratic programming problem yield the partial optimal estimator $\hat w_t(\beta_t)$ which is the eigenvector for $K(\beta_t)$ that correspond to the largest eigenvalue.

After defining the partial optimal estimators $\hat \beta_t(w_t), \hat w_t(\beta_t)$, we use the following n-step coordinate descent algorithm to estimate the global optimal $\tilde w_t$:

\begin{algorithm}[H]
\SetAlgoLined
\KwResult{Portfolio weight $\tilde w_t$}
 $w_t = w_{t}^{\text{Coint}} \quad \text{or } w_{t}^{MaxVar}$\;
 $\beta_t = \hat \beta_t(w_t) = \hat \beta_{t}^{OLS}$\;
 $step = 0$\;
 \While{step $< 1$}{
  $w_t = \hat w_t(\beta_t) = \hat w_{t}^{QP}$ \;
  $\beta_t = \hat \beta_t(w_t) = \hat \beta_{t}^{OLS}$\;
  $step+=1$ \;
 }
 $\tilde w_t = \frac{w_t}{|w_t|_1}$ \;
 \caption{VMAT(AR) n-step coordinate descent}
\end{algorithm}

$w_{t}^{\text{Coint}}, w_{t}^{MaxVar}$ denote the $w_t$ initiated by the cointegration vector or the vector that tries to maximize univariate series volatility. In practice, we usually only allow 1 iteration of the algorithm due to the fast convergence speed.

After we have determined the optimal $w_t$, we can develop $\delta_t$ by the same concept in the cointegration framework : cover the profiting probability. Recall that we define $y_t = w_t^T log X_t$.

\begin{align}
    \delta_{\alpha, t} = &\begin{cases}
    1 & y_t < \sigma^{Long}_{\alpha, t} \\
    0 & \text{else} \\
    -1 & y_t > \sigma^{Short}_{\alpha, t}
    \end{cases} \notag \\
    \text{where} &\quad \prod_{i=1}^d \Phi( \frac{\sigma^{Long}_{\alpha, t} - \hat y_{t+i}}{Err(\hat y_{t+i})}) = 1 - \alpha \notag \\
    &\quad \prod_{i=1}^d \Phi( \frac{\sigma^{Short}_{\alpha, t} - \hat y_{t+i}}{Err(\hat y_{t+i})}) = \alpha \notag
\end{align}

$\hat y_{t+i}$ denotes the $i$-step ahead forecast from $y_t$ by the fitted AR(p) model; $Err(\hat y_{t+i})$ denotes the forecast error of $\hat y_{t+l}$. $\Phi$ is the CDF function of the standard normal distribution.

As we mentioned in the last section, we use greedy approach for the bailing time $\tilde l_t$.

\subsection{Select optimal $\lambda$}

We develop two methods for the $\lambda$ selection: one is by the cross validation and the other is by the backward selection with goodness-of-fit tests.

For CV selection, we evaluate the VMAT strategy performance over several trading time before the current time with different choices of $\lambda$. Then we select the optimal $\lambda$ that has the highest averaged profit.

For the backward selection (later we name it VMAT Tame), we start the VMAT optimization with a large $\lambda$. We decrease the $\lambda$ until the residual of the AR(p) model doesn't reject the null hypothesis of the Ljung-Box tests.

The computational complexity of both methods are linear to the size of the search space in the worst case scenario. However, in practice, the second method is usually faster. Furthermore, it's easier to specify the search space of $\lambda$ in the second method.

%% file: section_3.tex
\section{Experiment}
\label{section_3}

In this section, we experiment our methods on two data sets and discuss multiple issues on them.

\subsection{Dataset}

We collect two data set for model evaluation:
\begin{itemize}
    \item Data 1: 5 years daily adjusted close price for Coca-Cola Co (KO) and PepsiCo Inc (PEP) from April 8th 2016 to April 9th 2021. The data is therefore a $(252*5) \times 2$ matrix.
    \item Data 2: 5 years daily adjusted close price for Apple Inc (AAPL), Amazon.com, Inc. (AMZN), Facebook, Inc. Common Stock (FB), Alphabet Inc Class C (GOOG), Netflix Inc (NFLX), Tesla Inc (TSLA), SPDR S\&P 500 ETF Trust (SPY) from April 8th 2016 to April 9th 2021. The data is therefore a $(252*5) \times 7$ matrix.
\end{itemize}

For either of the two dataset, we create three different data scenarios by specifying $d=3, 7, 14$, which represent strategies are allowed in short trading period, medium trading period and long trading period respectively. Intuitively, the shorter of the trading period is, the harder of profiting will be.

\subsection{Arbitrage Performance}
In this subsection, we study the arbitrage performance of different methods across different data scenarios.

\begin{itemize}
    \item Cointegration approach. It controls profiting probability based on the stationary distribution.
    \item Cointegration AR. It controls profiting probability based on fitting the AR model on the stationary process.
    \item MaxVar AR. It controls profiting probability by forming a maximized volatility univariate series, and fitting the AR model on it.
    \item VMAT. Our method with $\lambda = 1$.
    \item VMAT CV. Our method with $\lambda$ selected by cross validation.
    \item VMAT Tame. Our method with $\lambda$ selected by backward goodness-of-fit testing.
\end{itemize}

For all the methods above, we fix their parameters $p=10, L=60, \alpha=0.999$. The parameter values might not be the optimal in terms of empirical profiting performance. In later section, we also show that it's necessary for advanced fine tuning in order to achieve optimal profit. But for the simplicity of model comparison, we fixed them the same across different data scenarios.

The strategy of each method will be executed independently on each time point over the evaluation period. In the end, we collect profit and loss on each time point and compute the following performance metrics:
\begin{itemize}
    \item PL average. Sample mean and standard error will be reported.
    \item Signal rate. The rate of strategy participation (i.e. position not equals to zero).
    \item Control rate. The rate of strategy finishes arbitrage before reach the time limit $d$.
    \item Profit rate. The ratio of profiting days over all days. Non-participation days will count as non-profit days.
    \item Max draw. The maximum drawback (a.k.a the largest loss).
\end{itemize}

The results of scenarios "data 1, d = 7" and "data 2, d = 7" are shown below. The rest of tables are in \ref{appendix_1}.

\begin{table}[!h]
\centering
\caption{Data 1, $d = 7$. All values here are presented in percentage (i.e. x\%). 'SR' denotes the signal ratio. 'CR' denotes the control ratio. 'PR' denotes the profit ratio. 'maxDraw' denotes maximum drawback.}
\begin{tabular}{|c|cc|c|c|c|c|} 
    \hline
     & \multicolumn{2}{|c|}{PL mean (se)} & SR & CR & PR & maxDraw \\
    \hline
        Coint & 0.04549 & (0.02745) & 77.8 & 76.4 & 61.1 & -5.3\\
        Coint AR & 0.4785 & (0.01801) & \textbf{87.0} & \textbf{98.9} & \textbf{86.0} & \textbf{-2.6}\\
        MaxVar AR & 0.7366 & (0.03172) & 86.1 & 98.8 & 85.3 & -7.9\\
        VMAT & 0.7597 & (0.03278) & 85.2 & 98.7 & 84.4 & -7.1\\
        VMAT CV & 0.7619 & (0.03288) & 85.0 & 98.7 & 84.2 & -7.1\\
        VMAT Tame & \textbf{0.7687} & \textbf{(0.03281)} & 85.6 & 98.7 & 84.8 & -7.1\\
    \hline
\end{tabular}
\label{table_2}
\end{table}

\begin{table}[!h]
\centering
\caption{Data 2, $d = 7$. Values in \%.}
\begin{tabular}{|c|cc|c|c|c|c|} 
    \hline
     & \multicolumn{2}{|c|}{PL mean (se)} & SR & CR & PR & maxDraw \\
    \hline
        Coint & 0.01194 & (0.03170) & 87.9 & 75.6 & 68.4 & -12.\\
        Coint AR & 0.4090 & (0.01620) & \textbf{88.7} & \textbf{99.3} & 88.2 & \textbf{-0.6}\\
        MaxVar AR & 0.7779 & (0.03314) & 86.8 & 99.2 & 86.2 & -4.7\\
        VMAT & 1.02 & (0.04712) & 87.9 & 98.8 & 87.0 & -5.1\\
        VMAT CV & \textbf{1.07} & \textbf{(0.04888)} & 88.0 & 98.8 & \textbf{87.2} & -5.2\\
        VMAT Tame & 1.06 & (0.04819) & 88.1 & 98.8 & \textbf{87.2} & -5.2\\
    \hline
\end{tabular}
\label{table_5}
\end{table}

From the tables we can observe that in all data scenarios, VMAT with optimized $\lambda$ achieves the highest overall profit; On the other hand, VMAT exhibit higher profit when the dimension of multivariate series increases. This is an interesting phenomenon since traditional strategy usually fails at high dimension situation; Beside the profit, VMAT also achieves comparable signal ratio, control ratio and profiting ratio to the baseline methods. However, VMAT as a profiting strategy usually has higher max drawback comparing to the traditional cointegration approaches.

\subsection{Profit as a long-term strategy}
We plot the cumulative profit of VMAT over 5 years when $d = 7$. As it shows in \ref{appendix_2} Figure \ref{fig_1}, VMAT gains profit at most time points regardless of the fluctuation of the spread, especially when no systematic risks occurs; When systematic risks occurs, VMAT has the capability of turning it into profit. This can be observed by the surged profit around Jan 2020 when the recession in the stock market emerged.

\subsection{Sensitivity to parameters}
We test VMAT's performance over different values of the parameters. The experiments are performed on both data sets with fixed $d=3$. The results are shown in \ref{appendix_3} Figure \ref{fig_2}.

First of all, we test the impact of $\lambda$ on the list [1,3,5,7,10,13,20,30]. To our surprise, VMAT is resilient to different $\lambda$. For data 1, the profit is mostly unchanged; For data 2, no statistical difference in profit is found across different $\lambda$.

Then, we test the impact of different $\alpha$ on the list [0.4, 0.65, 0.8, 0.9, 0.95, 0.99]. The profit doesn't change much until alpha reaches a very large value.

We move on to test different $p$ with a list [5,7,10,13,17,21,25,30]. The lag order $p$ is influential to the profit as its presenting a upside down U-shape curve. Fine tuning is needed for $p$ in order to optimize the strategy profit.

At last, we test different $L$ in [30, 40, 50, 60, 70 ,80]. The pattern of profit over different $L$ seems different between the two data set: for data 2, the curve has U-shape while for data 1 it's the opposite. Since $L$ is also influential to the profit, fine tuning is needed in order to optimize the strategy profit. 

\subsection{Computation}
We study the behavior of the coordinate descent algorithm. We build VMAT strategy on the last possible trading time of two data sets with d = 3, L = 60, p = 10, $\lambda$ = 1. $w_1$ as the first element of $w$ is presented along iterations of the algorithm.

As shown from \ref{appendix_4} Figure \ref{fig_3} \ref{fig_4}, the algorithm is robust in initial method and converges by only one step.

%% file: section_4.tex
\section{Conclusion}
\label{section_4}

In this paper, we introduce an innovative statistical arbitrage framework for multivariate time series. It finds optimal portfolio weight by solving volatility \& model adaption trade-off, and determine participation by controlling the profiting probability in the trading period. Experiments have shown that VMAT has superior profit performance over the baselines including the traditional cointegration method especially when the length of trading period is limited.

%% file: appendix_1.tex
\section{Arbitrage performance}
\label{appendix_1}

\begin{table}[!h]
\caption{Data 1, $d = 3$.}
\centering
\begin{tabular}{|c|cc|c|c|c|c|} 
    \hline
     & \multicolumn{2}{|c|}{PL mean (se)} & SR & CR & PR & maxDraw \\
    \hline
        Coint & 0.004720 & (0.01656) & \textbf{33.5} & 49.6 & \textbf{20.3} & -3.5\\
        Coint AR & 0.08762 & (0.01404) & 13.8 & 72.8 & 10.9 & \textbf{-3.3}\\
        MaxVar AR & 0.1243 & (0.02669) & 12.0 & 74.3 & 9.85 & -10.\\
        VMAT & 0.1254 & (0.02663) & 11.8 & 76.7 & 9.85 & -10.\\
        VMAT CV & 0.1276 & (0.02692) & 12.4 & 76.5 & 10.2 & -10.\\
        VMAT Tame & \textbf{0.1297} & \textbf{(0.02698)} & 12.3 & \textbf{77.0} & 10.2 & -10.\\
    \hline
\end{tabular}
\label{table_1}
\end{table}

\begin{table}[!h]
\centering
\caption{Data 1, $d = 14$. Values in \%.}
\begin{tabular}{|c|cc|c|c|c|c|} 
    \hline
     & \multicolumn{2}{|c|}{PL mean (se)} & SR & CR & PR & maxDraw \\
    \hline
        Coint & 0.08717 & (0.03473) & 100. & 84.6 & 85.0 & -7.6\\
        Coint AR & 0.5345 & (0.01856) & 100. & 99.8 & 99.8 & \textbf{-2.2}\\
        MaxVar AR & 0.7709 & (0.03682) & 100. & 99.8 & 99.8 & -18.\\
        VMAT & 0.8031 & (0.03687) & 100. & 99.8 & 99.8 & -18.\\
        VMAT CV & 0.8073 & (0.03691) & 100. & 99.8 & 99.8 & -18.\\
        VMAT Tame & \textbf{0.8128} & \textbf{(0.03692)} & 100. & 99.8 & 99.8 & -18.\\
    \hline
\end{tabular}
\label{table_3}
\end{table}

\begin{table}[!h]
\centering
\caption{Data 2, $d = 3$. Values in \%.}
\begin{tabular}{|c|cc|c|c|c|c|} 
    \hline
     & \multicolumn{2}{|c|}{PL mean (se)} & SR & CR & PR & maxDraw \\
    \hline
        Coint & 0.02599 & (0.01930) & \textbf{50.0} & 50.4 & \textbf{32.3} & -9.1\\
        Coint AR & 0.06782 & (0.01113) & 13.9 & 77.2 & 11.6 & \textbf{-2.0}\\
        MaxVar AR & 0.1529 & (0.02430) & 14.0 & \textbf{86.3} & 12.6 & -8.4\\
        VMAT & 0.2063 & (0.03691) & 13.4 & 80.1 & 11.7 & -10.\\
        VMAT CV & \textbf{0.2276} & \textbf{(0.03858)} & 13.5 & 80.8 & 11.7 & -10.\\
        VMAT Tame & 0.2003 & (0.03754) & 13.5 & 79.6 & 11.9 & -10.\\
    \hline
\end{tabular}
\label{table_4}
\end{table}

\begin{table}[!h]
\centering
\caption{Data 2, $d = 14$. Values in \%.}
\begin{tabular}{|c|cc|c|c|c|c|} 
    \hline
     & \multicolumn{2}{|c|}{PL mean (se)} & SR & CR & PR & maxDraw \\
    \hline
        Coint & 0.0 & (0.04056) & 100. & 82.5 & 83.0 & -10.\\
        Coint AR & 0.4341 & (0.01661) & 100. & 99.9 & 99.9 & -0.9\\
        MaxVar AR& 0.8281 & (0.03119) & 100. & 99.8 & 99.8 & -3.4\\
        VMAT AR & 1.09 & (0.04444) & 100. & 99.9 & 99.9 & -3.0\\
        VMAT CV & \textbf{1.17} & \textbf{(0.04626)} & 100. & \textbf{100}. & \textbf{100}. & \textbf{0}\\
        VMAT Tame & 1.13 & (0.04640) & 100. & 99.9 & 99.9 & -2.6\\
    \hline
\end{tabular}
\label{table_6}
\end{table}

%% file: appendix_2.tex
\section{Long term performance}
\label{appendix_2}

\begin{figure}[!h]
    \centering
    \includegraphics[width=\textwidth]{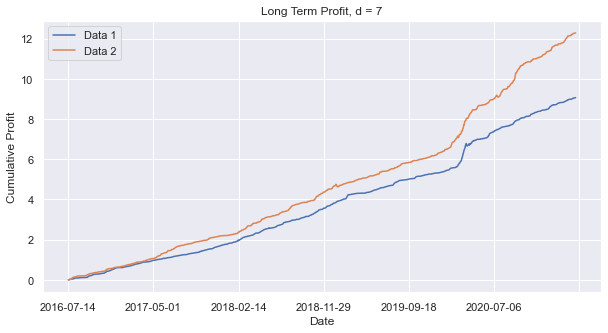}
    \caption{VMAT long term profit over 5 years. The values on y-axis is in normal scale.}
    \label{fig_1}
\end{figure}

%% file: appendix_3.tex
\section{Sensitivity to parameters}
\label{appendix_3}

\begin{figure}[!h]
    \centering
    \includegraphics[width=\textwidth]{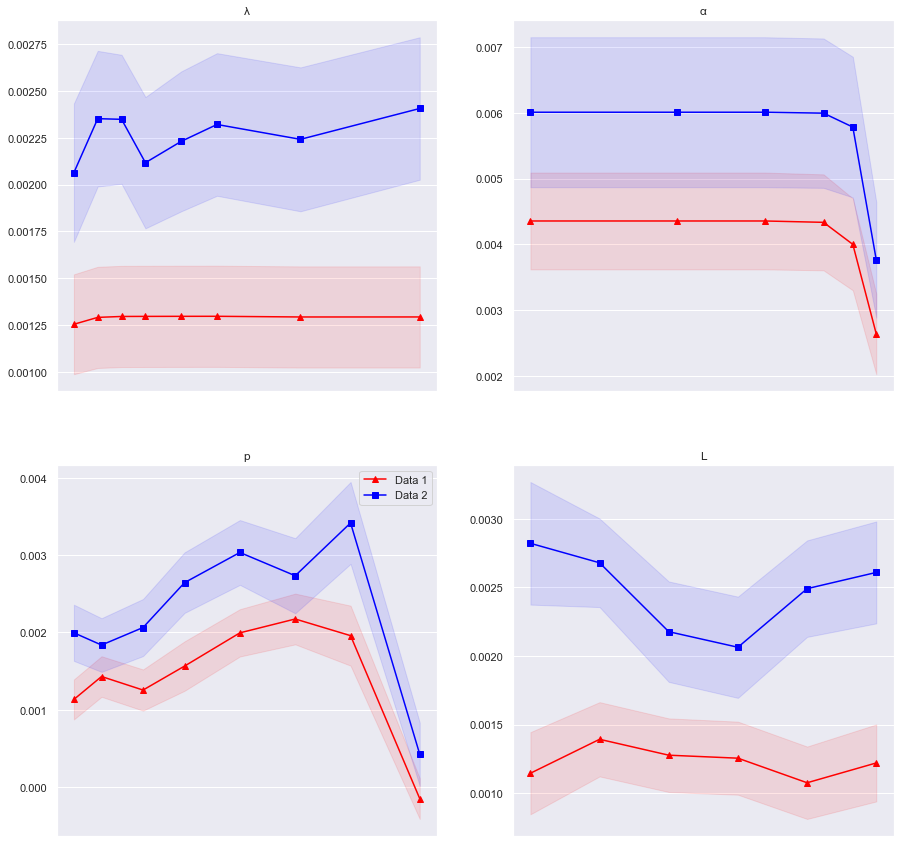}
    \caption{Profit against different parameters. The 95\% confidence interval is also presented.}
    \label{fig_2}
\end{figure}

%% file: appendix_4.tex
\section{Computation}
\label{appendix_4}

\begin{figure}[!h]
    \centering
    \includegraphics[width=\textwidth]{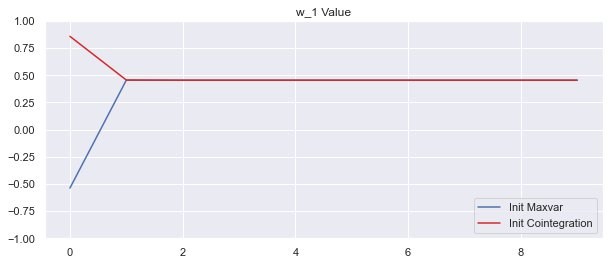}
    \caption{Convergence on data 1. 10 steps are evaluated. Algorithms with different initializing method are compared.}
    \label{fig_3}
\end{figure}

\begin{figure}[!h]
    \centering
    \includegraphics[width=\textwidth]{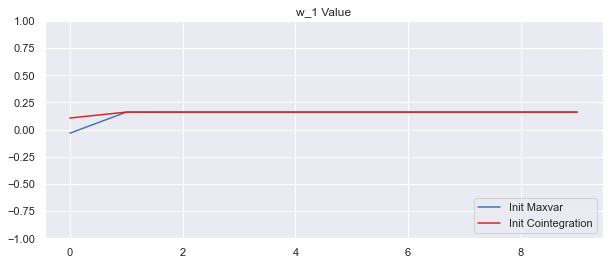}
    \caption{Convergence on data 2. 10 steps are evaluated. Algorithms with different initializing method are compared.}
    \label{fig_4}
\end{figure}